\documentclass[%
 aip, jap
 graphicx
 sd,%
 amsmath,amssymb,
 reprint,%
]{revtex4-1}

\usepackage{graphicx}
\usepackage{dcolumn}
\usepackage{bm}
\usepackage[utf8]{inputenc}
\usepackage[T1]{fontenc}
\usepackage[english]{babel}
\usepackage{color}

\newcommand{\beq}{\begin{equation}} 
\newcommand{\eeq}{\end{equation}}
\newcommand{\bfig}{\begin{figure}}
\newcommand{\efig}{\end{figure}}
\newcommand{\igraph}{\includegraphics}

\graphicspath {{C:/Users/caspe/Dropbox/Dottorato/memorie/resume/SUST-JAP-PRAdv/Immagini/}}

\begin{document}

\preprint{AIP}

\title{RF assisted switching in magnetic Josephson junctions}

\author{R. Caruso}
\affiliation{Dipartimento di Fisica, Università di Napoli Federico II; Monte Sant'Angelo - via Cintia; I-80126 Napoli - Italy}
\affiliation{CNR-SPIN, Monte S. Angelo - Via Cintia, I-80126 Napoli - Italy}
\author{D. Massarotti}
\affiliation{Dipartimento di Ingegneria Elettrica e delle Tecnologie dell'Informazione, Università di Napoli Federico II, Via Claudio, I-80125 Napoli, Italy}
\affiliation{CNR-SPIN, Monte S. Angelo - Via Cintia, I-80126 Napoli - Italy}
\author{V.V. Bol'ginov}
\affiliation{Institute of Solid State Physics (ISSP RAS), Chernogolovka, Moscow Region
142432 - Russia}
\affiliation{National University of Science and Technology MISIS, 4 Leninsky prosp., Moscow
119049 - Russia}
\author{A. Ben Hamida}
\affiliation{National University of Science and Technology MISIS, 4 Leninsky prosp., Moscow
119049 - Russia}
\affiliation{Leiden Institute of Physics, Leiden University, Niels Bohrweg 2, 2333 CA Leiden - The Netherlands}
\author{L.N. Karelina}
\affiliation{Institute of Solid State Physics (ISSP RAS), Chernogolovka, Moscow Region
142432 - Russia}
\author{A. Miano}
\affiliation{Dipartimento di Fisica, Università di Napoli Federico II; Monte Sant'Angelo - via Cintia; I-80126 Napoli - Italy}
\author{I.V. Vernik}
\affiliation{HYPRES, Inc. - 175 Clearbrook Road, Elmsford, NY 10523 - USA}
\author{F. Tafuri}
\affiliation{Dipartimento di Fisica, Università di Napoli Federico II; Monte Sant'Angelo - via Cintia; I-80126 Napoli - Italy}
\affiliation{CNR-SPIN, Monte S. Angelo - Via Cintia, I-80126 Napoli - Italy}
\author{V.V. Ryazanov}
\affiliation{Institute of Solid State Physics (ISSP RAS), Chernogolovka, Moscow Region
142432 - Russia}
\affiliation{Faculty of Physics, National Research University Higher School of Economics, Moscow - Russia
}
\author{O.A. Mukhanov}
\affiliation{HYPRES, Inc. - 175 Clearbrook Road, Elmsford, NY 10523 - USA}
\author{G.P. Pepe}
\affiliation{Dipartimento di Fisica, Università di Napoli Federico II; Monte Sant'Angelo - via Cintia; I-80126 Napoli - Italy}
\affiliation{CNR-SPIN, Monte S. Angelo - Via Cintia, I-80126 Napoli - Italy}

\date{\today}

\begin{abstract}

We test the effect of an external RF field on the switching processes of magnetic Josephson junctions (MJJs) suitable for the realization of fast, scalable cryogenic memories compatible with Single Flux Quantum logic. We show that the combined application of microwaves and magnetic field pulses can improve the performances of the device, increasing the separation between the critical current levels corresponding to logical '0' and '1'. The enhancement of the current level separation can be as high as 80\% using an optimal set of parameters. We demonstrate that external RF fields can be used as an additional tool to manipulate the memory states, and we expect that this approach may lead to the development of new methods of selecting MJJs and manipulating their states in memory arrays for various applications.

\end{abstract}



\maketitle

\section{Introduction}

Further progress in computing systems demands a significant improvement in energy-efficiency of digital data processing.  Cryogenic superconducting single flux quantum (SFQ) technology by virtue of its high speed and low power dissipation attracts a significant attention for the realization of the next generation classical computing system with high energy efficiency\cite{oleg1,oleg2}.  For progress in scalable fault-tolerant quantum computing, a cryogenic classical digital and memory technology is sought to provide supporting functions for qubit circuits such as readout, control, and error-correction. The realization of these functions with energy-efficient SFQ cryogenic technology provides an opportunity to locate these circuits in a close proximity to the superconducting qubit circuits \cite{oleg3, oleg4}.  While digital SFQ circuits have reached a relative maturity of a practical significance \cite{oleg5, oleg6, oleg7}, the matching low power dissipation, high density memory that could be integrated naturally with superconducting digital SFQ circuits remains elusive and still need to be demonstrated.  One of the problems is that despite significant efforts, the superconducting SFQ-based memories remain low density with relatively large size memory cells (tens of square microns) determined by large geometric inductances and transformers \cite{oleg8,oleg9}. This stimulates various hybrid memory approaches from the incorporation of entire semiconducting memory cell arrays \cite{oleg10} to the integration of spintronic memory elements \cite{oleg11, oleg12} and the development of memory devices based on magnetic Josephson junctions (MJJs) with ferromagnetic barriers \cite{oleg13, oleg14, oleg15, oleg16, oleg17}.  While achieving relative success in the memory device development, the overall designed memory cells remains large with their size dominated by elements such as nTrons \cite{oleg18}, readout SQUIDs, transformers required for enabling memory cell addressing within memory arrays, performing write and read operations.  Therefore, the development of alternative addressing approaches which would allow to decrease memory cell size and increase memory cell density will have a considerable impact on the cryogenic memory development.

In this paper, we study the effects of microwave fields on MJJ critical current levels in attempt to find an alternative addressing approach which would minimize the memory cell dimensions. 
The RF-assisted magnetization switching is a rather well known phenomenon for a wide range of systems such as magnetic clusters, single-domain magnetic particles and magnetic tunnel junctions \cite{RF1, RF2, RF3, RF4, RF5, RF6, RF7, RF8, RF9}, but it has never been investigated on MJJs, although spin wave resonances have been observed in conventional ferromagnetic Josephson junctions\cite{aprili}.
Here we show how this effect of remagnetization boost by RF fields can be used to improve discernibility of two logical states of a superconducting memory element based on Pd$_{0.99}$Fe$_{0.01}$ magnetic barrier.

When an RF field is applied together with a magnetic field pulse, the percentage difference in high and low critical current levels can be almost twice as large than the difference observed in absence of external RF fields.
We have found that for the particular sample investigated in this work the separation between the two current levels can be improved from 380 $\mu$A to 680 $\mu$A, for optimal working temperature and for magnetic field pulse amplitudes below 20\% of the saturation field.
This effect provides an additional knob to manipulate the memory state of MJJs which can increase power efficiency of the memory device and assist in solving MJJ addressing problems.

\section{Methods}

\bfig
\centering
\igraph{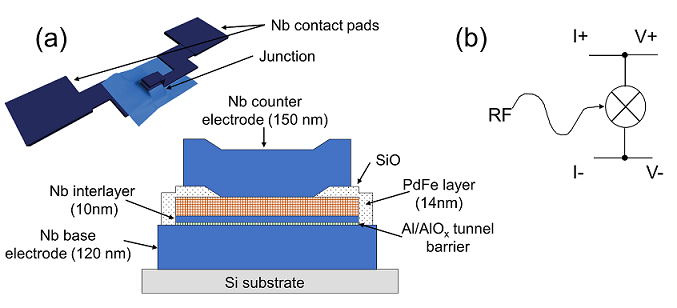}
\caption{The sample analyzed in this work: (a) MJJ 3D structure and cross section (not to scale) with layers thicknesses marked and (b) measurement wiring with I- and V- connected to the base electrode and I+ and V+ connected to the counter electrode.}
\label{fig:sketch}
\efig

The MJJs used for this experiment have been fabricated within a collaboration between HYPRES Inc. and the Russian Institute of Solid State Physics (ISSP-RAS)\cite{oleg16}. The first step of the fabrication is the production of a Nb/Al-AlO$_x$(Nb) trilayer using the standard HYPRES process to attain 4.5 kA/cm$^2$ critical current density samples \cite{hypres_std, hypres_std2}. The bottom Nb electrode is 120 nm, while the Nb counter electrode is 15 nm thick at this stage. Then, the wafers are diced in 15 mm x 15 mm samples and sent to ISSP for the second stage of the fabrication. The Nb counter electrode is etched down to about 10 nm using Ar ion milling. This operation removes Nb oxide and possible organic residues from the Nb surface, providing good interface transparency between the Nb layer and ferromagnetic barrier. Then, a Pd$_{0.99}$Fe$_{0.01}$/Nb bilayer is deposited on top using rf- and dc-magnetron sputtering in Ar plasma. The thickness of the PdFe layer is 14 nm and the top Nb layer is 150 nm thick. 
Further fabrication process corresponds to that described elsewhere\cite{oleg16, Josephson_magnetometry}. The junction mesa has a square shape of 10x10 $\mu$m$^2$ size. A sketch of the sample is shown in Fig. \ref{fig:sketch}a, while the measurement wiring is shown in Fig. \ref{fig:sketch}b. The important point noted in previous works\cite{oleg17, SIsFS} is that superconductivity is not completely suppressed in the intermediate Nb layer, allowing the transmission of a sufficiently large supercurrent through this SIsFS  multilayer as compared to a SIFS junction, where S is for superconductor, I for insulator and F for ferromagnet. On the other hand, thickness of the Nb interlayer is much smaller than London penetration depth, and any magnetic effect due to external field or internal magnetization affects simultaneously tunnel and ferromagnetic layers. This ensures both hysteretic critical current versus magnetic field dependence $I_C (H)$ due to remagnetization of PdFe barrier and high $I_C R_N$ product due to tunnel barrier of the SIsFS junction, where $I_C$ is the critical current of the junction and $R_N$ the normal state resistance. In other words, experimental data show that AlO$_x$-(Nb)-PdFe multilayer serves as a single high-resistive barrier with hysteretic remagnetization curve.
\begin{figure}[tbp]
\centering
\includegraphics{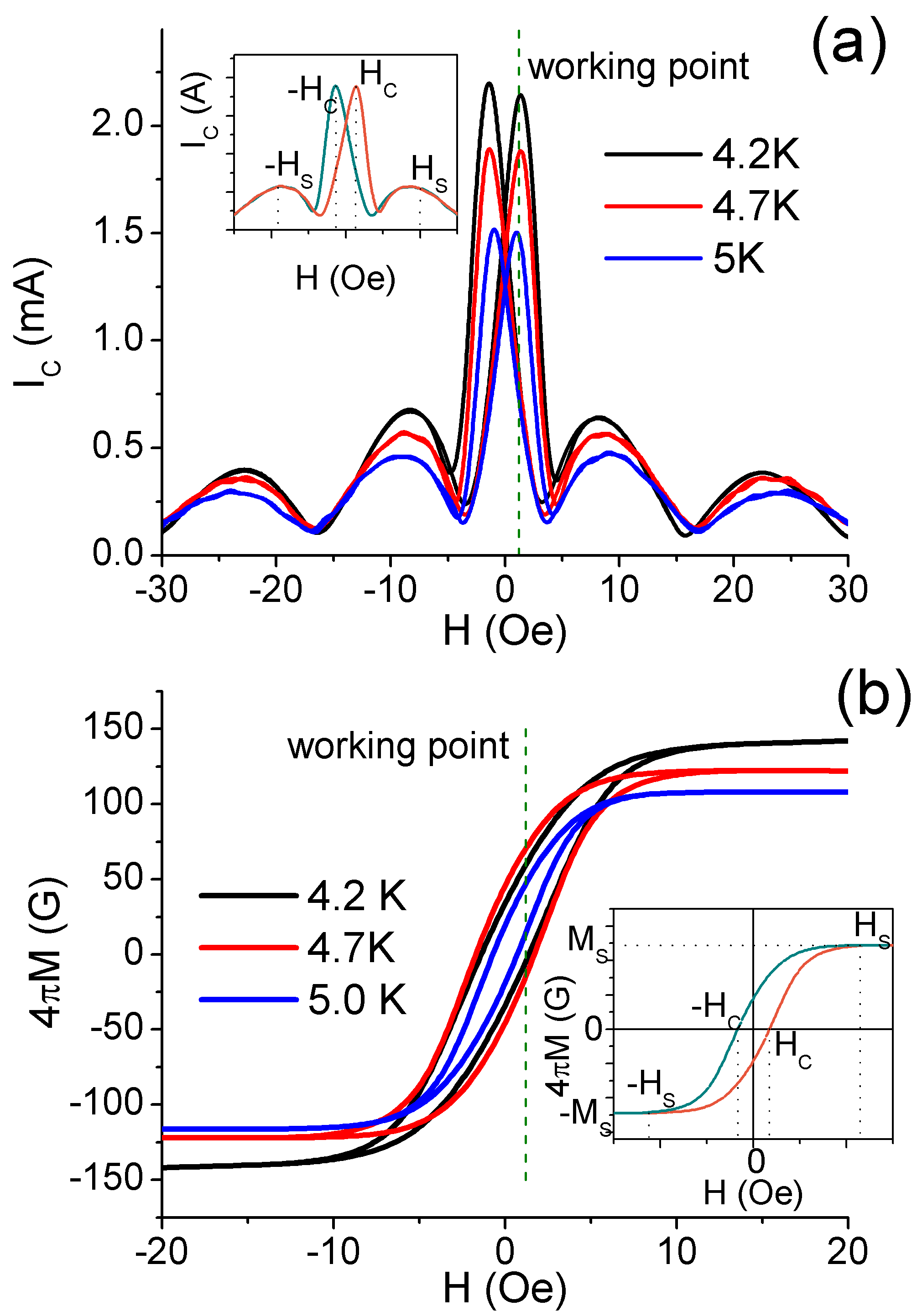}
\caption{Magnetic field dependence of the critical current (a) and of the magnetization (b) for a MJJ with PdFe thickness of 14 nm at different temperatures. Green line represent the chosen working point. As temperature increases, the spacing between the two maxima of I$_C$(H) decreases (a). Similarly, the width of M(H) curves decreases as T increases (b). Insets: characteristic fields of the ferromagnet (coercive field $H_C$ and saturation field $H_S$ indicated on sample curves).}
\label{fig:ICHeMH}
\end{figure}

The measurements have been performed in a Heliox VL evaporation cryostat, with the sample thermally anchored to the $^3$He pot of the cryostat. Magnetic field is applied in the plane of the junction, which is the easy magnetization plane due to the presence of the two superconducting electrodes acting as screens, using a superconducting Nb/Ti coil thermally anchored to the 1 K pot of the cryostat to ensure thermal insulation of the sample. We use a pulse generator to control width, amplitude and other parameters such as delay, rising and falling time of the pulse. The filtering system of our cryostat is designed to minimize thermal and electronic noise, using a combination of RC filters and copper powder filters anchored to different thermal stages, with cutoff frequencies of about 1 MHz and 1 GHz respectively\cite{PRB2011}. Standard current-voltage characteristics (I-Vs)  have been measured as a function of temperature and magnetic field, by current biasing the MJJ with a ramp at low frequency (about 10 Hz) and by measuring the voltage across the junction with a battery powered differential amplifier. The RF train is emitted by an antenna placed close to the sample and controlled through a microwave generator synchronized with the signal generator used for magnetic field pulses, so that it is possible to control its delay with respect to the magnetic field pulse and its length, as well as its frequency and power.

The working frequency is 3.88 GHz, chosen so that the coupling with the junction is maximum, which means that we observe the maximum critical current reduction when microwaves are continuously applied to the sample. This frequency is slightly higher than the ferromagnetic resonance (FMR) frequencies detected for 100 nm thick Pd$_{0.99}$Fe$_{0.01}$ layers \cite{RF8}, and provides effective magnetization dynamics due to microwaves. The exact optimal microwave frequency we use for this experiment only depends upon the geometry of our setup, i.e. the relative positions of the microwave antenna and the sample holder. At other frequencies, the coupling between the antenna and the sample is strongly reduced. 
All the measurements presented in this work have been performed setting the RF generator power level to 4.9 dBm and the train duration to 250 ms at 4.2 K, if not specified otherwise. 
The power level of the microwave field is then reduced by two 3 dB attenuators anchored at different stages of the cryostat.
Measurements performed at base temperature (0.3 K) on samples with lower critical current confirm the results obtained at 4.2 K. The working temperature has been chosen as it is the typical working temperature of Nb-based SFQ circuits.

The use of an antenna limits our control of the microwave train, and cannot be considered as a viable approach for memory cell fabrication. However, the demonstration of RF influence on memory switching processes paves the way for the development of new addressing schemes based on coplanar RF waveguides, as shown for magnetic tunnel junctions\cite{RF2}.

PdFe layer is a soft ferromagnet with cluster magnetic structure and a Curie temperature of about 10 K at 10-20 nm thickness\cite{RF8,PRBgol}. The magnetic moment of Pd$_{0.99}$Fe$_{0.01}$ thin layers is distributed mostly within the relatively large Fe-rich Pd$_3$Fe nanoclusters of approximately 10 nm size and around 100 nm spacing in between. The clusters are embedded in a paramagnetic Pd host layer with high polarizability\cite{magn}. The hysteresis loop is due to reorientation of the magnetic moment of these clusters.

In this work, we choose Pd$_{0.99}$Fe$_{0.01}$ because its properties have been extensively studied \cite{RF8, PRBgol, magn}. 
The excitation of magnetic moments out of their equilibrium position is not limited to a specific material, so our results on Pd$_{0.99}$Fe$_{0.01}$ can be extended to other ferromagnetic materials. More specifically, it is possible to use single-domain ferromagnetic interlayers with biaxial magnetization anisotropy such as CrO$_2$\cite{eschrig} or Permalloy\cite{perm1}, where the RF signal could ease the switch between the two corresponding energy minima. Another possibility is the use of PdFe alloys with different Fe concentrations. The ferromagnetic properties of the alloy can be tuned using an appropriate Fe concentration, thus improving the scalability of the memory element.

Since the magnetization is in the plane of the junction, the magnetic flux $\Phi_M$ due to the layer magnetization $M$ adds up to the flux generated by the external field $\Phi_H$
\beq
\Phi=\Phi_M+\Phi_H=4\pi M L d_F + H L d_m
\label{eqn:phi}
\eeq
where $L$ is the junction width, $d_F$ is the thickness of the ferromagnetic layer and $d_m$ is the magnetic thickness of the SIsFS junction\cite{oleg17, Josephson_magnetometry, SIsFS}. The $I_C(H)$ dependence of the junction is thus a shifted and distorted Fraunhofer pattern (Fig. \ref{fig:ICHeMH}a) consistently with theory\cite{Josephson_magnetometry, SIsFS}. For each value of the magnetic field below the saturation field, there are two different critical current values that can be used as logic states ('1' and '0'). $M(H)$ curves can be calculated from $I_C(H)$ curves (Fig. \ref{fig:ICHeMH}) using the known methods \cite{Josephson_magnetometry}. 
%
%

From I$_C$(H) it is also possible to estimate experimentally $d_m$, the magnetic thickness of the SIsFS junction. For fields above the saturation field, the magnetic flux $\Phi_M$ due to magnetization is constant, and thus the difference between two subsequent minima depends solely on the applied field\cite{Josephson_magnetometry}:
\beq
\Delta\left(\frac{\Phi}{\Phi_0}\right)= \Delta\left(\frac{\Phi_H}{\Phi_0}\right)=\Delta H L d_m
\label{eqn:magn_thick}
\eeq
From $M(H)$ curves it is possible to estimate the saturation magnetization $M_S$, the coercive field $H_C$ and the saturation field $H_S$ (inset in Fig.\ref{fig:ICHeMH}a and Fig.\ref{fig:ICHeMH}b). $M_S$ is the maximum magnetization value reached by the ferromagnetic barrier, at 3.5 K we observe $M_S \approx$ 150 G, while at 5 K we measure a saturation magnetization around 100 G. $H_C$ is the external field needed to completely demagnetize the ferromagnet, and in our case ranges from $\approx$ 1.2 Oe at 3.5 K  to 0.8 Oe at 5 K. The saturation field $H_S$ is the field at which $M_S$ is reached. For our samples it is almost constant between 3.5 K and 4.7 K, approximately 10 Oe, and it drops to roughly 6 Oe at 5.0 K.

For uniform MJJ\cite{Josephson_magnetometry, SIsFS} the two $I_C(H)$ branches are symmetrical with respect to zero field, and so the memory cell use of this device requires a magnetic field bias to set the optimal working point. This corresponds to the magnetic field value for which the difference between the high and the low critical current levels is as large as possible, provided that the magnetic bias is within the saturation field. This occurs in correspondence of the maximum of $I_C(H)$ pattern, at about 1.2 Oe, as shown in Fig. \ref{fig:ICHeMH}.
In previous works \cite{oleg16} this was achieved automatically due to magnetic field generated by bias current.

\section{Results}

We compare the current levels obtained applying only magnetic field pulses 500 ms long (Fig. \ref{fig:pan1}a, left) and those obtained applying microwaves together with the field pulse. In both cases critical current levels have been measured from standard I-V curves acquired after the end of magnetic field pulse. In Fig. \ref{fig:pan1}a (right), the RF train has been modulated so that it is 250 ms long and centered around the center of the field pulse (Fig. \ref{fig:pan1}a, right). The time width of the train is tunable, along with the delay with respect to the magnetic field pulse and the rise and fall parameters. The comparison has been performed collecting 10 pairs of low-high current I-Vs for each case, and $\Delta I$ is calculated from the average of low and high current levels using Eq. \ref{eqn:deltaI}. Uncertainty on $\Delta I$ is estimated by propagating the errors on the average high and low current levels.

\bfig[htbp]
\centering
\igraph{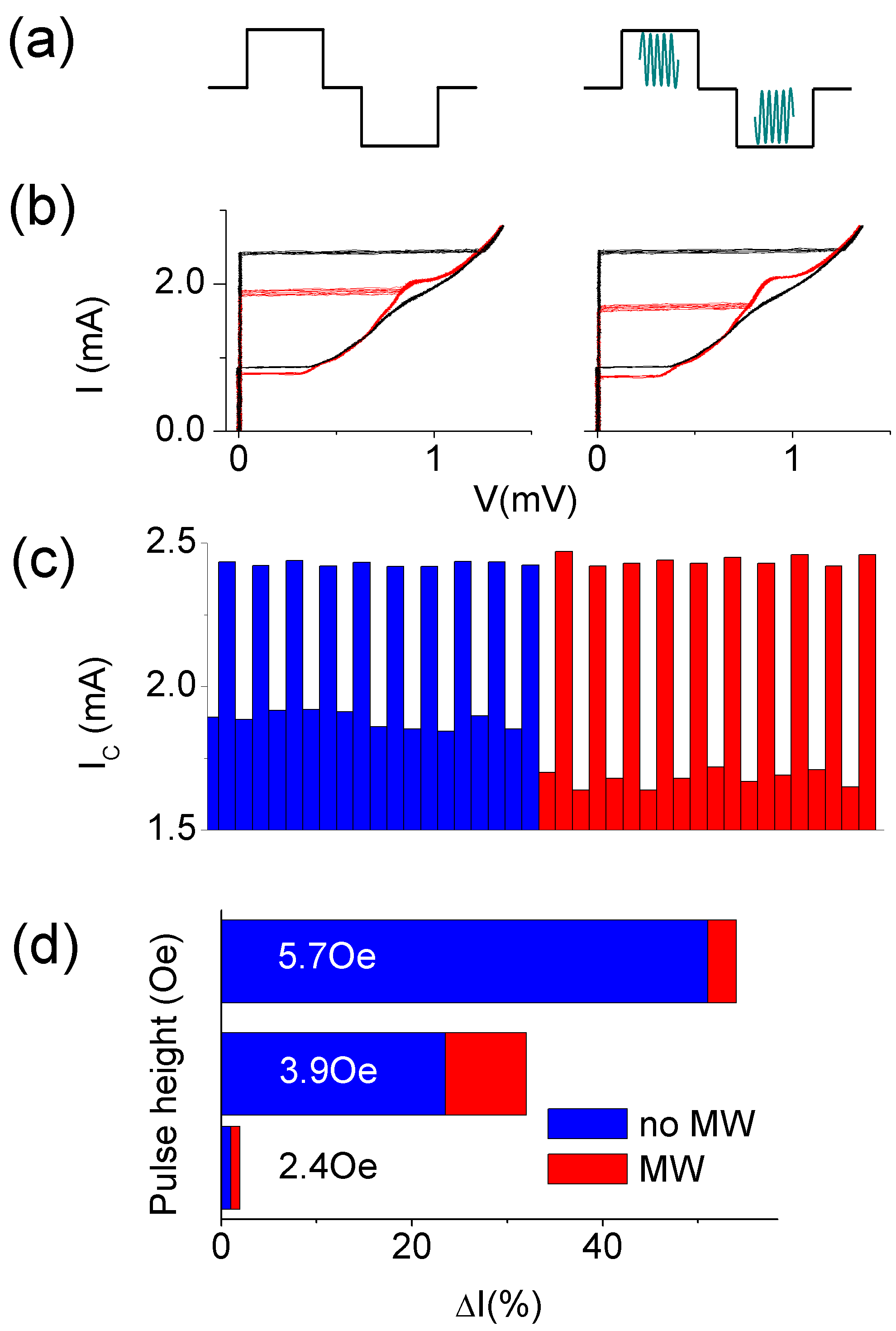}
\caption{(a) Scheme of the driving pulses. Black: magnetic field pulse, green: microwaves. (b) and (c) IV curves and current levels in absence (left) and in presence (right) of microwaves. The amplitude of the field pulse is 3.9 Oe, the RF train has a frequency of 3.88 GHz and a duration of 250 ms. (b) Black IV curves are registered at high critical current level, red IVs at low critical current levels. (c) Blue bars represent current levels in absence of applied microwaves, red bars are current levels when an RF field is applied together with magnetic field pulses. (d) Comparison between $\Delta I$ with and without microwaves for different values of the field pulse amplitude. Blue: $\Delta I$ when switching is performed using only magnetic field pulses, red: $\Delta I$ when microwaves are used to assist the switching process.}
\label{fig:pan1}
\efig 

An example of the above mentioned procedure is shown in Fig. \ref{fig:pan1}b where we present I-V curves for high $I_C$ states (black) and low $I_C$ states (red). On the left we show I-V curves when only magnetic field pulses are applied, on the right I-V curves registered after the application of magnetic field and microwaves (see Fig. \ref{fig:pan1}a (right)) are shown. The column plot in Fig. \ref{fig:pan1}c shows the critical current levels obtained from each I-V curve in Fig. \ref{fig:pan1}b. Blue bars represent the current levels in absence of microwaves, while the red bars are the current levels when microwaves are applied together with field pulses.

Firstly we tested the devices as memory elements using large magnetic field pulses (i.e. with amplitude larger than the saturation field) as reported in literature\cite{oleg16}, and then we applied the microwave train. To evaluate the effect we define $\Delta I$ as the percentage difference in high and low critical current levels, and $G$ as the percentage enhancement of $\Delta I$:
\begin{eqnarray}\label{eqn:deltaI}
\Delta I&=&\frac{I_C^{high}-I_C^{low}}{I_C^{high}} \\ 
G&=&\frac{\Delta I_{MW}-\Delta I_{noMW}}{\Delta I_{noMW}}
\label{eqn:G}
\end{eqnarray}

For these field values, the difference between critical current levels $\Delta I$ is (60 $\pm$ 2)\% in both cases, so $G$ is zero. Decreasing the field pulse amplitude down to $\approx H_S/2 \approx$ 5.7 Oe, $\Delta I$ becomes higher when the RF field is applied, and $G$ increases. By further decreasing the field pulse amplitude, $G$ increases and reaches the maximum difference at about 3.9 Oe, and then decreases progressively, until at low fields, of about 2.4 Oe, $G$ goes back to zero. As shown in Fig. \ref{fig:pan1}d, we observed an enhancement of $\Delta I$ in a wide range of field pulses. For a field pulse of 3.9 Oe, the percentage enhancement $G$ is 35\%.

\bfig
\centering
\igraph{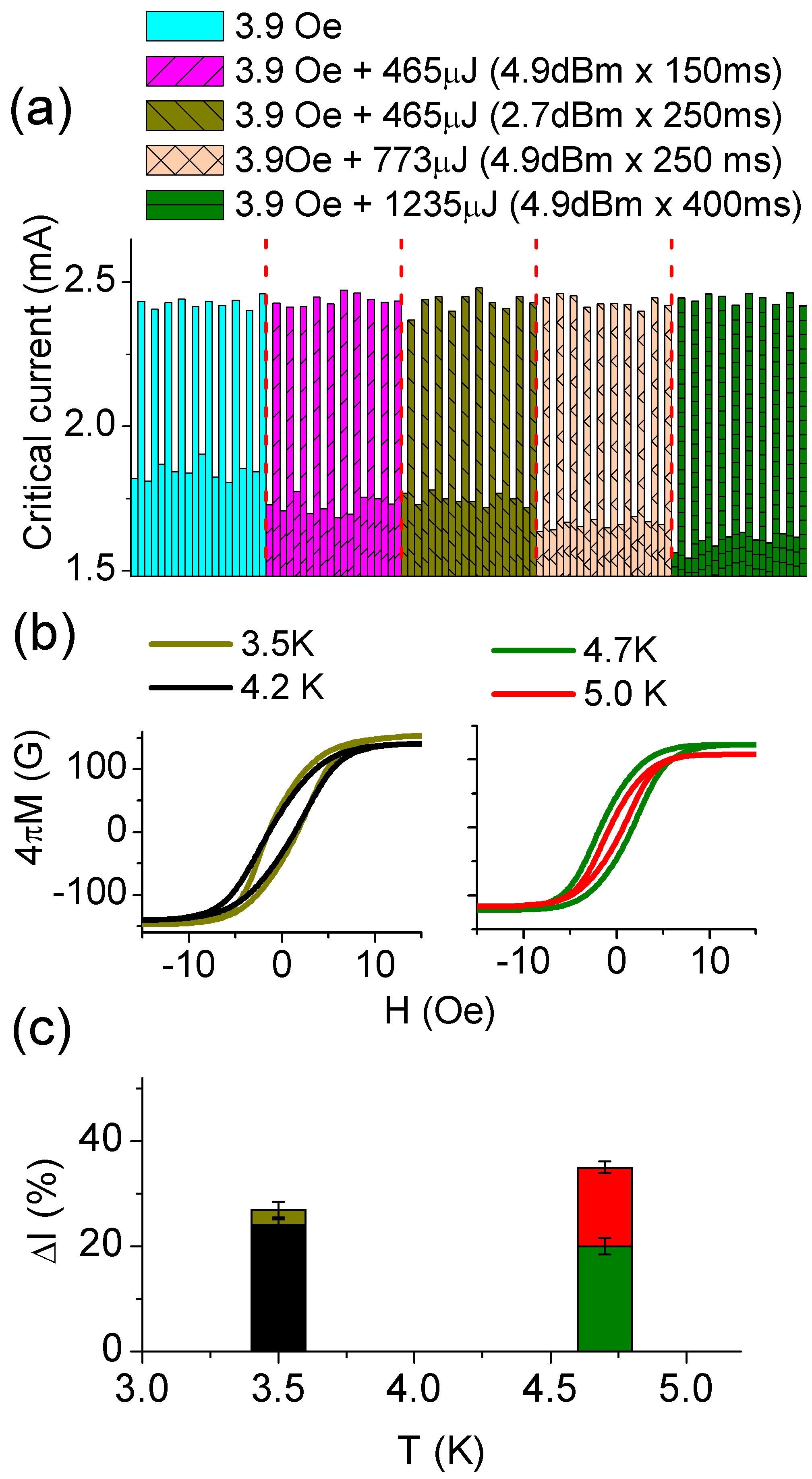}
\caption{(a) Current levels for different duration and power level of the microwave train. The working temperature is 4.2 K, while the pulse amplitude is 3.9 Oe. (b) Magnetization curves obtained from I$_C$(H) measurements at different temperatures. Left: 3.5 K and 4.2 K. Right: 4.7 K and 5 K (c) $\Delta I$ for different working temperatures. Assuming a local heating $\approx 0.5 K$, the difference in $\Delta I$ obtained with and without microwaves is justified by the difference in M(H) curves shown in panel (b).}
\label{fig:varE}
\efig

\subsection{Energy and temperature dependence}

Our experiments show that the difference between the critical current levels also depends on the amount of energy transferred to the device.
In particular, we tested the effect for different RF power levels and pulse durations, alongside with testing at different nominal energy values of the microwaves. We have found that increasing RF power level and keeping the train duration constant, the current level separation enhancement increases, and the same happens when increasing the time duration of the external microwave field keeping the power level constant\cite{noiTAS}. 

In Fig. \ref{fig:varE}a we show the current level separation obtained for three nominal energy values, 465 $\mu$J, 773 $\mu$J and 1235 $\mu$J. For the lowest energy value, we show current levels obtained for different power and time duration values of the microwave train, chosen so that the energy is kept almost constant. In this case the current levels separation is the same within error bars. The comparison between all the five data-sets in Fig. \ref{fig:varE}a confirms that the power level or the duration alone play no active role in the switching enhancement.

The use of long pulses for magnetic fields and RF trains is due to intrinsic limits of the setup used to demonstrate the effect. The characteristic time of the superconducting coil used to generate magnetic field is of the order of several ms, while the use of an antenna to apply microwaves implies a reduction of the actual power reaching the sample. In order to measure the intrinsic switching times of the MJJ, a further step is needed to implement a specific setup and engineer a sample equipped with dedicated on-chip waveguides.

The dependence of critical current levels on the energy carried by the RF signal demonstrates an instability in cluster magnetic structure of PdFe induced by the RF signal itself.
The microwave radiation excites fluctuations of local magnetic moments of PdFe clusters that decrease the coercive field and facilitate the remagnetization process. 
The alternating field of the microwave signal moves the local magnetic moments out of equilibrium, causing their precession, which can be treated as thermal fluctuations. 
This analogy is suggested by temperature measurements shown in Fig. \ref{fig:varE}b and Fig. \ref{fig:varE}c.

The percentage enhancement $G$ at a given temperature corresponds to the percentage difference in coercive fields at different temperatures.
At low temperatures, M(H) is almost constant, so if the effective temperature of local magnetic moments varies from the 3.5 K working temperature to 4.2 K due to the effect of microwaves the coercive field $H_C$ and the saturation magnetization $M_S$ are almost unchanged, as sketched in Fig. \ref{fig:varE}b (left). Therefore the difference between $\Delta I_{MW}$ and $\Delta I_{no MW}$ is small, as shown in Fig. \ref{fig:varE}c (left), where the corresponding $G$ is 8\%. At higher temperatures (for instance for a working temperature of 4.7 K), $M(H,T)$ and $M(H,T+\delta T)$ can have significantly different coercive fields, as can be seen from Fig. \ref{fig:varE}b (right), and thus a higher difference between $\Delta I$ with and without applied microwaves is observed in Fig. \ref{fig:varE}c (right), where the percentage difference between $\Delta I_{MW}$ and $\Delta I_{no MW}$ is as large as 80\%.
This enhancement proves that microwave and electrical signals can be used together to write information to a MJJ. This creates other design possibilities for memory array addressing schemes.

\section{Discussion}

The change in M(H) curves around 4.7 K has been recently explained \cite{magn} taking into account a two-component magnetization of PdFe. 
It has been shown that thin (< 20 nm) PdFe films with low iron content, present two different Curie temperatures that correspond to two different interactions. The main contribution is due to short range interaction, while the weaker contribution is related to a Ruderman-Kittel-Kasuya-Yosida (RKKY) long range interaction.  In our samples, PdFe is 14 nm thick, with a higher Curie temperature of $\sim$12 K. According to previous works\cite{magn}, the lower Curie temperature $T_1$ is $\sim 0.4 T_2$ for 20 nm PdFe thin films, where $T_2$ is the higher Curie temperature. In our case, $T_1$ corresponds to $\sim$4.8 K, which is in agreement with our experimental data.  

Another potential application of the MJJ with PdFe-nanoclustered ferromagnetic layer is in neuromorphic computing circuits. Recently, it was proposed to use MJJs with Mn ferromagnetic nanoclusters for the implementation of artificial synapses compatible with SFQ-based artificial neurons as underlying technology platform for large scale neuromorphic systems\cite{olegnew1,olegnew2}. One can use microwave radiation to tune critical current of MJJ-based synapses implementing synaptic plasticity – the foundation of learning and memory functions in neuromorphic systems.

The proposed microwave-assisted tuning of MJJ critical currents may be used also in the recently proposed superconducting magnetic field-programmable gate array (FPGA), in which MJJs are used to implement the FPGA programming layer\cite{olegnew3}. 
 
The MJJ device with the clustered ferromagnetic layer may not be scalable to submicron dimensions, perhaps limiting their size to ~1 $\mu$m$^2$.  However, the size of the MJJ is often not defining the cryogenic memory cell dimensions, nor the memory array density.  In fact, the addressing schemes to enable a reliable selection of memory cell in a random access memory (RAM) array define the memory cell area and RAM array density.  In particular, the few square micron MJJ used in JMRAM\cite{olegnew4} is embedded into a readout SQUID which dominates the memory cell area and may not be scalable below ~20 $\mu$m$^2$. This makes the perspective MJJ-based memory cell comparable in size to the all-JJ memory cell\cite{oleg9} re-scaled using modern fabrication processes.

Most of the literature devoted to new cryogenic memory devices does not address this issue and focuses solely on the memorizing storage element such as MJJs of various kinds overlooking the critical issue of implementing the addressable memory cell design within a 2D RAM array configuration.  Therefore, we believe that the development of new ways to address and manipulate memory cell within the array will have the highest impact on the eventual success of various memory implementations achieving higher density.  This work on microwave assisted MJJ switching is a step in this direction.

\section{Conclusion}

In conclusion, we observed a clear evidence of RF-assisted switching of magnetic Josephson junctions in a wide range of magnetic field pulses. For field pulses above and below this range, the RF radiation does not provide any effect.  The larger field pulses magnetize MJJ into a saturation and any change in M(H) induced by microwaves cannot be distinguished.  For the lower field pulses, M(H) becomes non-hysteretic and almost linear. The RF-assisted switching mechanism can be explained in terms of effective heating of local magnetic moments, and its temperature dependence can be interpreted in terms of a two-component magnetization in PdFe. The application of RF radiation makes possible to induce MJJ switching using lower field pulses than without RF.  This also allows one to reduce reading bias current, potentially leading to the improvement of the overall energy efficiency. 

The damping of the coercive field caused by the microwave signal can be used to choose the amplitude of the writing magnetic pulse in such a way that only MJJs subjected to the microwave radiation change their digital state. In this way, the addressing task is reduced to the control of the microwave train, which is easier to perform. Further studies are required to determine the role of other mechanisms that might be involved in switching processes. 

We plan to extend the RF-assisted switching approach to entirely RF addressing of memory cells in an array, using coplanar waveguides as described elsewhere\cite{RF1,RF2}, which can allow us to better tune other parameters that can improve the overall performances of the memory array.

\begin{acknowledgments}
This work has been partially financed by \emph{B5 2F17 001400005} grant within the framework of SEED 2017 of CNR-SPIN. V.V. Bolginov and L.N. Karelina thank Russian Foundation for Basic Research grant \emph{17-02-01270}, A. Ben Hamida acknowledges the Ministry of Education and Science of the Russian Federation in the framework of Increase Competitiveness Program of NUST "MISiS" (research project N.\emph{K4-2014-080}), V.V. Ryazanov thanks joint Russian-Greece grant \emph{2017-14-588-0007-011}. The authors would also like to thank \emph{NANOCOHYBRI} project (Cost Action CA 16218).
\end{acknowledgments}


\begin{thebibliography}{0}%
\makeatletter
\providecommand \@ifxundefined [1]{%
 \@ifx{#1\undefined}
}%
\providecommand \@ifnum [1]{%
 \ifnum #1\expandafter \@firstoftwo
 \else \expandafter \@secondoftwo
 \fi
}%
\providecommand \@ifx [1]{%
 \ifx #1\expandafter \@firstoftwo
 \else \expandafter \@secondoftwo
 \fi
}%
\providecommand \natexlab [1]{#1}%
\providecommand \enquote  [1]{``#1''}%
\providecommand \bibnamefont  [1]{#1}%
\providecommand \bibfnamefont [1]{#1}%
\providecommand \citenamefont [1]{#1}%
\providecommand \href@noop [0]{\@secondoftwo}%
\providecommand \href [0]{\begingroup \@sanitize@url \@href}%
\providecommand \@href[1]{\@@startlink{#1}\@@href}%
\providecommand \@@href[1]{\endgroup#1\@@endlink}%
\providecommand \@sanitize@url [0]{\catcode `\\12\catcode `\$12\catcode
  `\&12\catcode `\#12\catcode `\^12\catcode `\_12\catcode `\%12\relax}%
\providecommand \@@startlink[1]{}%
\providecommand \@@endlink[0]{}%
\providecommand \url  [0]{\begingroup\@sanitize@url \@url }%
\providecommand \@url [1]{\endgroup\@href {#1}{\urlprefix }}%
\providecommand \urlprefix  [0]{URL }%
\providecommand \Eprint [0]{\href }%
\providecommand \doibase [0]{http://dx.doi.org/}%
\providecommand \selectlanguage [0]{\@gobble}%
\providecommand \bibinfo  [0]{\@secondoftwo}%
\providecommand \bibfield  [0]{\@secondoftwo}%
\providecommand \translation [1]{[#1]}%
\providecommand \BibitemOpen [0]{}%
\providecommand \bibitemStop [0]{}%
\providecommand \bibitemNoStop [0]{.\EOS\space}%
\providecommand \EOS [0]{\spacefactor3000\relax}%
\providecommand \BibitemShut  [1]{\csname bibitem#1\endcsname}%
\let\auto@bib@innerbib\@empty
\end{thebibliography}%


\begin{thebibliography}{99}

\bibitem{oleg1}	D. S. Holmes, A. L. Ripple, M. A. Manheimer, \emph{IEEE Trans. Appl. Supercond.}, \textbf{23}, 1701610 (2013).

\bibitem{oleg2}	O. A. Mukhanov, \emph{IEEE Trans. Appl. Supercond.}, \textbf{21}, 760 (2011).

\bibitem{oleg3} R. McDermott and M. G. Vavilov, \emph{Phys. Rev. Applied} \textbf{2}, 014007 (2014).

\bibitem{oleg4}	K. G. Fedorov, A. V. Shcherbakova, M. J. Wolf, D. Beckmann, and A. V. Ustinov, \emph{Phys. Rev. Lett} \textbf{112}, 160502 (2014).

\bibitem{oleg5}	Y. Ando, R. Sato, M. Tanaka, K. Takagi, N. Takagi, A. Fujimaki, \emph{IEEE Trans. Appl. Supercond.}, \textbf{26}, 1301205 (2016).

\bibitem{oleg6} O.A. Mukhanov,, D. Kirichenko, I.V. Vernik, T.V. Filippov, A. Kirichenko, R. Webber, V. Dotsenko, A. Talalaevskii, J.C. Tang, A. Sahu, P. Shevchenko, R. Miller, S.B. Kaplan, S.Sarwana and D. Gupta, \emph{IEICE Trans. Electron.}, \textbf{E91-C(3)}, 306 (2008)

\bibitem{oleg7} T. V. Filippov, A. Sahu, A. F. Kirichenko, et al., \emph{Physics Procedia} \textbf{36}, 59 (2012).

\bibitem{oleg8}	P. F. Yuh, \emph{IEEE Trans. Appl. Supercond.} \textbf{3} 3013 (1993).

\bibitem{oleg9} S. Nagasawa, H. Numata, Y. Hashimoto, and S. Tahara, \emph{IEEE Trans. Appl. Supercond.} \textbf{9} 3708 (1999).

\bibitem{oleg10} T. Van Duzer, L. Zheng, S. R. Whiteley, H. Kim, J. Kim, X. Meng, T. Ortlepp \emph{IEEE Trans. Appl. Supercond.}, \textbf{23}, 1700504 (2013).

\bibitem{oleg11} L. Ye, D. B. Gopman, L. Rehm, D. Backes, G. Wolf, T. Ohki, A.F. Kirichenko, I.V. Vernik, O.A. Mukhanov, A.D. Kent, \emph{J. Appl. Phys.}, \textbf{115}, 17C725 (2014).

\bibitem{oleg12} S. V. Aradhya, G. E. Rowlands, J. Oh, D. C. Ralph, and R. A. Burman, \emph{Nano Lett.}, 2016.

\bibitem{oleg13} B. Baek, W. H. Rippard, S. P. Benz, S. E. Russek, P. D. Dresselhaus, \emph{Nat. Commun.}, \textbf{5} 3888 (2014).

\bibitem{oleg14} M. A. E. Qader, R. K. Singh, S. N. Galvin, L. Yu, J.M. Rowell, N. Newman \emph{Appl. Phys. Lett.}, \textbf{104}, 022602 (2014).

\bibitem{oleg15} B. M. Niedzielski, E. C. Gingrich, R. Loloee, W.P. Pratt, N.O. Birge , \emph{Supercond. Sci. Technol.} \textbf{28}, 085012 (2015).

\bibitem{oleg16} T.I. Larkin, V.V. Bol'ginov, V.S. Stolyarov, V.V. Ryazanov, I.V. Vernik, S.K. Tolpygo, and O.A. Mukhanov, \emph{Appl. Phys. Lett.}, \textbf{100}, 222601 (2012).

\bibitem{oleg17} I. V. Vernik, V. V. Bol'ginov, S. V. Bakurskiy, A.A. Golubov, M. Yu. Kupriyanov, V.V: Ryazanov, O.A. Mukhanov, \emph{IEEE Trans. Appl. Supercond.} \textbf{23} 1701208, (2013).

\bibitem{oleg18} A. N. McCaughan and K. K. Berggren, \emph{Nano Letters} \textbf{14}, 5748, (2014)

\bibitem{RF1} C. Thirion, W. Wernsdorfer, and D. Mailly \emph{Nat. Materials} \textbf{2} 524 (2003).

\bibitem{RF2} T. Moriyama, R. Cao, J. Q. Xiao, X. R. Wang, Q. Wen, and H. W. Zhang \emph{Appl. Phys. Lett.} \textbf{90} 152503 (2007).

\bibitem{RF3} C. Raufast, A. Tamion, E. Bernstein, V. Dupuis, T. Tournier, T. Crozes, E. Bonet, Edgar and W. Wernsdorfer \emph{IEEE Trans. Magn.} \textbf{44} 2812 (2008).

\bibitem{RF4} X. Fan, Y. S. Gui, A. Wirthmann, G. Williams, D. Xue, and C.-M. Hu, \emph{Appl. Phys. Lett.} \textbf{95} 062511 (2009).

\bibitem{RF5} S. Okamoto, N. Kikuchi, O. Kitakami, T. Shimatsu, and H. Aoi \emph{J. Appl. Phys.} \textbf{109} 07B748 (2011).

\bibitem{RF6} L. Cai, D. A. Garanin, and E. M. Chudnovsky \emph{Phys. Rev. B} \textbf{87} 024418 (2013).

\bibitem{RF7} H. Suto, T. Kanao, T. Nagasawa, K. Kudo, K. Mizushima, and R. Sato \emph{Appl. Phys. Lett.} \textbf{110} 262403 (2017).

\bibitem{RF8} I. A. Golovchanskiy, V. V. Bolginov, N. N. Abramov, V. S. Stolyarov, A. Ben Hamida, V. I. Chichkov, D. Roditchev, and V. V. Ryazanov \emph{J. Appl. Phys.} \textbf{120} 163902 (2016).

\bibitem{RF9} S. J. Graves, H. Muraoka, and Y. Kanai \emph{AIP Advances} \textbf{7} 056517 (2017).

\bibitem{aprili} I. Petkovic, M. Aprili, S. E. Barnes, F. Beuneu and S. Maekawa, \emph{Phys. Rev. B} \textbf{80} 220502(R) (2009).

\bibitem{hypres_std} D. Yohannes, S. Sarwana, S.K. Tolpygo, A. Sahu, and V. Semenov \emph{IEEE Trans. Appl. Supercond.} \textbf{15} 90 (2005).

\bibitem{hypres_std2} S.K. Tolpygo, D. Yohannes, R.T. Hunt, J.A. Vuvalda, D. Donnelly, D. Amparo and A.F. Kirichenko \emph{IEEE Trans. Appl. Supercond} \textbf{17} 946 (2007).

\bibitem{Josephson_magnetometry} V.V. Bol'ginov, V.S. Stolyarov, D.S. Sobanin, A.L. Karpovich, and V.V. Ryazanov \emph{JETP Lett.} \textbf{95}, 366 (2012).

\bibitem{SIsFS} S.V. Bakurskiy, N.V. Klenov, I.I. Soloviev, V.V. Bol'ginov, V.V. Ryazanov,
I.V. Vernik, O.A. Mukhanov, M.Yu. Kupriyanov, and A.A. Golubov \emph{Appl. Phys. Lett} \textbf{102}, 192603 (2013).

\bibitem{PRB2011} L. Longobardi, D. Massarotti, G. Rotoli, D. Stornaiuolo, G. Papari, A. Kawakami, G. P. Pepe, A. Barone, F. Tafuri \emph{Appl. Phys. Lett.} \textbf{99} 062510 (2011).

\bibitem{PRBgol} I. A. Golovchanskiy, V. V. Bol’ginov, V. S. Stolyarov, N. N. Abramov, A. Ben Hamida, O. V. Emelyanova, V. S. Stolyarov, M. Yu. Kupriyanov, A. A. Golubov, and V. V. Ryazanov \emph{Phys. Rev. B} \textbf{94} 214514 (2016).

\bibitem{magn} V.V. Bol'ginov, O.A. Tikhomirov, L.S. Uspenskaya \emph{JETP Lett.} \textbf{105}, 169 (2017).


\bibitem{eschrig} M. Eschrig and T. L\"ofwander, \emph{Nat. Phys.}, \textbf{4}, 138 (2008).

\bibitem{perm1} I.D. Bursuc, R. Grimberg and E. Diaconu \emph{Physics Letters A} \textbf{50}, 359 (1974).

\bibitem{noiTAS} R. Caruso, D. Massarotti, A. Miano, V. V. Bol'ginov, A. Ben Hamida, L. N. Karelina, G. Campagnano, I. V. Vernik, F. Tafuri, V. V. Ryazanov, O. A. Mukhanov and G. P. Pepe \emph{IEEE Trans. Appl. Supercond., submitted for publication.} 

\bibitem{olegnew1} S. E. Russek, C. A. Donnelly, M. L. Schneider, B. Baek, M. R. Pufall, W. H. Rippard, and P. F. Hopkins \emph{2016 IEEE International Conference on Rebooting Computing (ICRC)}, pp. 1-5, (2016)

\bibitem{olegnew2} M. L. Schneider, C. A. Donnelly, S. E. Russek, B. Baek, M. R. Pufall, P. F. Hopkins, W. H. Rippard, \emph{2016 IEEE International Conference on Rebooting Computing (ICRC)}, pp. 1-4 (2017)

\bibitem{olegnew3} N. Katam, O. Mukhanov, M. Pedram, "Superconducting magnetic field programmable gate array", \emph{IEEE Trans. Appl. Supercond., 2018, in press.}

\bibitem{olegnew4} I. M. Dayton, T. Sage, E. C. Gingrich, M. G. Loving, T. F. Ambrose, N. P. Siwak, S. Keebaugh, C. Kirby, D. L. Miller, A. Y. Herr, Q. P. Herr, O. Naaman \emph{arXiv:1711.01681v1} (2017)

\end{thebibliography}
\end{document}